\documentclass[prl,twocolumn]{revtex4}
\usepackage{amssymb}
\usepackage{amsmath}

\input{tcilatex}

\begin{document}

\title{Characterizing the process of reaching consensus for social systems}
\author{ Jinn-Wen Wu$^{1}$, Yu-Pin Luo$^{2}$, and Ming-Chang Huang$^{3}$}
\email{ming@phys.cycu.edu.tw}
\affiliation{${\ }$ ${\ }^{1}$Department of Applied Mathematics, Chung-Yuan Christian
University, Chungli, 32023 Taiwan\\
$^{2}$Department of Electronic Engineering, National Formosa University,
Huwei, 63201, Taiwan\\
$^{3}$Department of Physics, Chung-Yuan Christian University, Chungli, 32023
Taiwan }

\begin{abstract}
A novel way of characterizing the process of reaching consensus for a social
system is given. The foundation of the characterization is based on the
theorem which states that the sufficient and necessary condition for a
system to reach the state of consensus is the occurrence of \textit{%
communicators}, defined as the units\ of the system that can directly
communicate with all the others simultaneously. A model is proposed to
illustrate the characterization explicitly. The existence of \textit{%
communicators} provides an efficient way for unifying two systems that a
state of consensus is guaranteed after the mergence.
\end{abstract}

\maketitle


\label{s0}

The appearance of system-wide harmonic behaviors, such as the globally
coordinated movements for the units of a system, the consensus of opinions
for a public event in a society, and etc., can be observed very often for
different systems in different situations\cite%
{lampl,boyd,krapivsky,young,bikhchandani}. It is very remarkable that there
is no center control in the first place for the occurrence of such global
coordinations. Then, it should be interested to know the kinematic scenario
for the arising of coordinated behaviors. In this Letter, we intend to give
a novel way of characterizing the process of reaching the state of consensus
for social systems. The cornerstone for the characterization is the
identification of \textit{communicators}, defined as the units of a system
that can directly reach all the others at some instant in the time evolution
of a system. Different units may start to act as \textit{communicators} at
different times, but, the units have remained to be in the same status once
they become \textit{communicators}. Then, we can classify the \textit{%
communicators}\ into different levels according to their first appearance
times. The \textit{primary} \textit{communicators} are referred to those
appearing in the earliest, and they may correspond to the hub-units which
are those with large values of degree in a social network. As the
distribution of hub-units has a strong effect on the scaling behavior of the
relaxation time towards the state of system-wide coordination from a
strongly disorder one\cite{zhou}, we will show that the presence of \textit{%
communicators} is the sufficient and necessary condition for a system to
achieve the state of consensus. Thence,\ the process of reaching the state
of consensus can be viewed as the sequential appearance of \textit{%
communicators}. By employing the Watts-Strogatz networks for the social
connections, we propose a simple model for the transition matrix to
illustrate the sequential pattern explicitly. The model study indicates that
the \textit{communicators} may appear in different levels except the case of
regular\ Watts-Strogatz network for which, all the \textit{communicators}
are \textit{primary}, and the \textit{communicators} with larger values for
the degree of connection, in general, appear earlier. The characterization
may provide useful applications to different situations. An example of
applications, the mergence of two groups, is given. As two groups, which
have the respective state of consensus, are merged together, our
characterization can provide an efficient way of combination that guarantee
the existence of a reachable state of consensus for the combined system. \ 

Consider a system of $N$ units with the communication paths specified by the
connection edges. The distribution of edges is given by a $N\times N$
connection matrix $\Gamma $ with the entries given as $\gamma _{ij}=1$ for
the existence of a directional edge from unit $j$ to $i$, and $\gamma
_{ij}=0 $ otherwise.\ An attitude-variable, denoted as $x_{i}$ for the unit $%
i$ with the value in the range $\left[ 0,1\right] $, is assigned to an unit
to represent its degree of favor towards an event; the attitude can be
viewed as complete disagreement for the value $0$, neutrality for $1/2$, and
complete agreement for $1$. As all variables take the same value, $x_{i}=c$
for $c\in \left[ 0,1\right] $ and $i=1$, $2$, $\cdot \cdot \cdot $, $N$, the
system is said to be in the state of consensus $c$. The time evolutions of
the variables $x_{i}$ in discrete time-steps are given as 
\begin{equation}
X\left( t+1\right) =M\cdot X\left( t\right) ,  \label{eqa-1}
\end{equation}%
where $M$ is the transition matrix and $X\left( t\right) =\left( x_{1}\left(
t\right) ,x_{2}\left( t\right) ,\cdot \cdot \cdot ,x_{N}\left( t\right)
\right) ^{\tau }$ with the superscript $\tau $ for the transpose. The
off-diagonal entry $m_{ij}$ of the matrix $M$ gives the fractional rate of
the influence from unit $j$ to unit $i$, and the diagonal entries $m_{ii}$
defines the fractional rate of persistence on the $x_{i}$ value. The
explicit form of $M$ depends on the model, but we assume that the feedback
from the self-persistence and the environmental influence is positive, this
renders the matrix $M$ to be non-negative..Furthermore, the fractional rates
are normalized, $\sum_{j=1}^{N}m_{ij}=1$ for $i=1$, $2$, $\cdot \cdot \cdot $%
, $N$, then, the $x_{i}$ values always lie in the range $\left[ 0,1\right] $
during the course of time evolution. These constraints for the matrix $M$
make the transpose of $M$, $M^{\tau }$, to be a stochastic matrix.

As $M$ is a non-negative square matrix, the Perron-Frobenius theorem can be
employed to assert the properties of the leading eigenvalue and the
corresponding eigenvectors\cite{berman}. Here, we establish a theorem, which
gives less restriction on the entries of $M$ than the Perron-Frobenius
theorem and provides the central theme for characterizing the process of
reaching the state of consensus in a system. As the theorem followed by its
proof are given below, here we summarize the notation conventions and the
corresponding definitions used in the theorem and the proof. $\left(
i\right) $ A bar is placed over the head of a vector, say $\overline{X}$, to
denote a stochastic vector associated with the stochastic matrix $M^{\tau }$%
, a stochastic vector is subject to the conditions that the components are
positive and the sum of the components equals $1$. $\left( ii\right) $
Different types of norm for a vector are used for convenience: $\parallel
X\parallel _{\infty }$is the super-norm of the vector $X$, defined as $%
\left\Vert X\right\Vert _{\infty }=\max \left\{ \left\vert x_{i}\right\vert
,0\leq i\leq N\right\} $; $\left\Vert X\right\Vert _{1}$ is the one-norm,
defined as $\left\Vert X\right\Vert _{1}=\sum_{i=1}^{N}\left\vert
x_{i}\right\vert $; and $\left\Vert X\right\Vert _{2}$ is the two-norm,
defined as $\left\Vert X\right\Vert _{2}=\left( \sum_{i}x_{i}^{2}\right)
^{1/2}$. Different types of norms are equivalent. $\left( iii\right) $\ The
bracket of two vectors, $\left\langle X,Y\right\rangle
=\sum_{i=1}^{N}x_{i}y_{i}$, denotes the inner product of $X$ and $Y$. \ \ \ 

\textit{Theorem}: Suppose there exists an unit, say $\alpha $, which can
connect every other unit by a path of length $n_{0}$, that is, the entries
of the $\alpha $-th column of $M^{n_{0}}$ are positive. Then, every
trajectory of solution for Eq. (\ref{eqa-1}) is leaded to a state of
consensus, 
\begin{equation}
X\left( k\right) \rightarrow cI\text{ as }k\rightarrow \infty .
\label{eqa-2}
\end{equation}%
Here, $I$ denotes the column vector with each entry $1$, and the $c$ value,
which signifies the state of consensus, is given as 
\begin{equation}
c=\left\langle X\left( 0\right) ,\overline{\Lambda }\right\rangle ,
\label{eqa-2b}
\end{equation}%
where $X\left( 0\right) $ is the initial state of the system, and $\overline{%
\Lambda }$ is the eigenvector of $M^{\tau }$ with eigenvalue $1$, $M^{\tau
}\cdot \overline{\Lambda }=\overline{\Lambda }$. Here, the existence and the
uniqueness of $\overline{\Lambda }$ are guaranteed by the Perron-Frobenius
theorem\cite{berman}. For the speed of convergence, there exists $r\geq 1$
and $0<\lambda <1$ such that 
\begin{equation}
\parallel X\left( k\right) -cI\parallel _{\infty }\leq r\lambda
^{k}\parallel X\left( 0\right) -cI\parallel _{\infty }.  \label{eqa-2c}
\end{equation}%
Moreover, the condition for leading to Eq. (\ref{eqa-2}) is also necessary.

We first give the proof for the sufficient condition by showing the
equivalent form of Eq. (\ref{eqa-2c}), 
\begin{equation}
\left\Vert X\left( k\right) -cI\right\Vert _{1}\leq r\lambda ^{\left[ k/n_{0}%
\right] }\left\Vert X\left( 0\right) -cI\right\Vert _{1},  \label{eqb-1}
\end{equation}%
held for the system evolving to the time step $k$ with $k\gg n_{0}$, where $%
\left[ k/n_{0}\right] $ is the integer part of $k/n_{0}$. To show the
inequality of Eq. (\ref{eqb-1}), we consider the dynamics of the stochastic
matrix $M^{\tau }$, 
\begin{equation}
\overline{Y}\left( k\right) =M^{\tau }\cdot \overline{Y}\left( k-1\right) .
\label{eqa-3}
\end{equation}%
As $M^{n_{0}}$ has a positive column, the matrix $\left( M^{\tau }\right)
^{n_{0}}$ has a positive row. By defining 
\begin{equation}
\tau =\sum_{i=1}^{N}\min \left\{ \left[ \left( M^{\tau }\right) ^{n_{0}}%
\right] _{ij}\text{, }1\leq j\leq N\right\} ,  \label{eqa-4}
\end{equation}%
we have $0<\tau <1$, this yields $0<\lambda <1$ for $\lambda =1-\tau $.
Following the theorem shown the Appendix of Ref. \cite{huang}, we have 
\begin{equation}
\left\Vert \left( M^{\tau }\right) ^{k}\left( \overline{Y}\left( 0\right) -%
\overline{Z}\left( 0\right) \right) \right\Vert _{1}\leq \lambda ^{\left[
k/n_{0}\right] }\left\Vert \overline{Y}\left( 0\right) -\overline{Z}\left(
0\right) \right\Vert _{1},  \label{eqa-5}
\end{equation}%
for $k\gg n_{0}$, where $\overline{Y}\left( 0\right) $ and $\overline{Z}%
\left( 0\right) $ are two different initial states for the dynamics of Eq. (%
\ref{eqa-3}). Furthermore, because of $M^{\tau }\cdot \overline{\Lambda }=%
\overline{\Lambda }$, we have 
\begin{equation}
\left\langle M^{n_{0}}\cdot \left( X\left( 0\right) -cI\right) ,\overline{%
\Lambda }\right\rangle =0,  \label{eqa-6}
\end{equation}%
with $c$ given by Eq. (\ref{eqa-2b}). Then, Eq. (\ref{eqb-1}) is followed
from Eqs. (\ref{eqa-5}) and (\ref{eqa-6}). To see this, we first notice that 
\begin{equation}
\left\Vert M^{k}\cdot X\left( 0\right) -cI\right\Vert _{1}=\left\Vert
M^{k}\cdot \left( X\left( 0\right) -cI\right) \right\Vert _{1}.
\label{eqa-6b}
\end{equation}%
Then, based on Eqs. (\ref{eqa-6}) and (\ref{eqa-6b}) we have 
\begin{equation}
\left\Vert M^{k}\cdot X\left( 0\right) -cI\right\Vert _{1}=\sum_{\alpha
=1}^{N}\left\vert \left\langle X\left( 0\right) -cI,\left( M^{\tau }\right)
^{k}\cdot \left( I_{\alpha }-\overline{\Lambda }\right) \right\rangle
\right\vert ,  \label{eqa-7a}
\end{equation}%
where $I_{\alpha }$ is a column vector with only one non-zero entry of the
value $1$ locating at the $\alpha $-th row, that is, $\sum_{\alpha
=1}^{N}I_{\alpha }=I$. The Cauchy inequality is further applied to the right
hand side of Eq. (\ref{eqa-7a}) to obtain 
\begin{equation}
\left\Vert X\left( k\right) -cI\right\Vert _{1}\leq K_{2}^{2}\left\Vert
X\left( 0\right) -cI\right\Vert _{1}\left\{ \sum_{\alpha =1}^{N}\left\Vert
\left( M^{\tau }\right) ^{k}\cdot \left( I_{\alpha }-\overline{\Lambda }%
\right) \right\Vert _{1}\right\} ,  \label{eqa-8}
\end{equation}%
where $K_{2}$ is the constant of equivalence between the two-norm and
one-norm. Finally, we apply Eq. (\ref{eqa-5}) to the second factor on the
right hand side of (\ref{eqa-8}) to obtain 
\begin{equation}
\left\Vert X\left( k\right) -cI\right\Vert _{1}\leq 2\cdot N\cdot
K_{2}^{2}\cdot \lambda ^{\left[ k/n_{0}\right] }\left\Vert X\left( 0\right)
-cI\right\Vert _{1},  \label{eqa-9}
\end{equation}%
where we use the fact, $\left\Vert I_{\alpha }-\overline{\Lambda }%
\right\Vert _{1}\leq \left\Vert I_{\alpha }\right\Vert _{1}+\left\Vert 
\overline{\Lambda }\right\Vert _{1}=2$. \ 

\ For the proof of necessary condition, we set the initial state as $X\left(
0\right) =\sum_{\alpha =1}^{N}I_{\alpha }$ and assume that the state of
consensus is reached as 
\begin{equation}
X\left( k\right) =M^{k}\cdot X\left( 0\right) \rightarrow cI  \label{eqa-10}
\end{equation}%
for $k\rightarrow \infty $, where the $c$ value is $c=\sum_{\alpha
=1}^{N}c_{\alpha }$ with $c_{\alpha }$ corresponding to the value for the
state of consensus when the initial state is $I_{\alpha }$. By observing
that the $\alpha $-th column of $M^{k}$ is $M^{k}\cdot I_{\alpha }$, we
write 
\begin{equation}
M^{k}=\left[ M^{k}\cdot I_{1},M^{k}\cdot I_{2},\cdot \cdot \cdot ,M^{k}\cdot
I_{N}\right] .  \label{eqa-11}
\end{equation}%
Then, based on Eq. (\ref{eqa-10}) we have 
\begin{equation}
M^{k}\rightarrow \left[ c_{1}I,c_{2}I,\cdot \cdot \cdot ,c_{N}I\right]
\label{eqa-12}
\end{equation}%
for sufficiently large $k$. Suppose that all $c_{\alpha }=0$ for $\alpha
=1,2,\cdot \cdot \cdot ,N$, this contradicts with the fact that $\rho \left(
M\right) =\rho \left( M^{\tau }\right) =1$, where $\rho \left( M\right) $ is
the spectrum radius of $M$. Thus, there is at least one $c_{\alpha }\neq 0$
in Eq. (\ref{eqa-12}), and this gives the condition in the \textit{theorem}
as necessary. \ \ 

Based on the \textit{theorem}, we define the unit $\alpha $ as a \textit{%
communicator} with the first appearance time $n_{0}$, if $n_{0}$ is the
smallest one among all integers $n$ that the $\alpha $-th column is positive
for the matrix $M^{n}$. Since the $\alpha $-th column remains to be positive
for $M^{n}$ with $n>n_{0}$ if it is positive for $M^{n_{0}}$, an unit once
become a \textit{communicator}, it remains to be a \textit{communicator}
afterwards. Thus, we can classify the \textit{commutators} into the \textit{%
primary}, the \textit{secondary}, and etc. in the order of the first
appearance time from the earliest to the latest, and characterize the
process of reaching a state of consensus by the sequential appearance of 
\textit{communicators}. However, the \textit{theorem} does not imply that
all units have to become \textit{commutators} before reaching the state of
consensus, although this may occur for some forms of $M$. \ \ \ 

For the purpose of illustration, we consider a simple model for the
transition matrix $M$. The fractional rate for the persistence of the
present attitude is assumed to be the same for all units and given by the
parameter $s$ with the setting $m_{ii}=s$, where $0\leq s<1$ and $%
i=1,2,\cdot \cdot \cdot ,N$. For the off-diagonal entries, we assume that
the environmental influence comes from the connected neighbors given by the
connection matrix $\Gamma $. Moreover, the average of the attitudes of the
neighbors is used to represent the social atmosphere faced by an unit. These
amount to set the off-diagonal entries as $m_{ij}=\left( 1-s\right) \gamma
_{ij}/z_{i}$, where $\gamma _{ij}$ are the entries of $\Gamma $, and $%
z_{i}=\tsum_{j=1}^{N}\gamma _{ij}$ is the inward degree of the unit $i$. The
undirected Watts-Strogatz networks are used to define the connection
matrices $\Gamma $. We first place $N$ units around a circle with the degree
of an unit $k_{0}$ connecting to the right and to the left neighbors
symmetrically; then a value, called rewiring probability $p$, is assigned to
rewire the edges randomly\cite{watts}. Consequently, the members of
Watts-Strogatz networks have different degrees of randomness from regular
lattices $\left( p=0\right) $ to random graphs $\left( p=1\right) $. For a
symmetric $\Gamma $, the matrix $M$ is symmetric and stochastic. Then, the
eigenvector of the eigenvalue $1$ for $M^{\tau }$ is $\overline{\Lambda }%
=\left( 1/N\right) I$, this leads to the state of consensus as the mean
value of the initial state, $c=\tsum_{i=1}^{N}x_{i}\left( 0\right) /N$,
which gives the state of consensus $c=1/2$ for a strongly disorder initial
state. By setting set $N=1000$, $k_{0}=4$, and $p=0.1$ for the network and $%
s=0.3$ for the self-persistence, we show the results in Fig. 1(a) for the
first appearance time $t_{c}$ of a \textit{communicator} $n$ in a trajectory
from a strongly disorder state to the state of consensus $c=1/2$ (the upper
part) and the corresponding degree of connection $k$ of the \textit{%
communicator} $n$ (the lower part). The results indicate that there does not
exist a definite relation between the first appearance time of a \textit{%
communicator} and its degree of connection. However, the corresponding $k$
values, in general, are larger for the \textit{communicators} that the $%
t_{c} $ values are smaller as shown in Fig. 1(b), where, based on the
results of Fig. 1(a), the average of the first appearance time of the 
\textit{communicators} with the same $k$ value, $\left\langle
t_{c}\right\rangle $, as a function of $k$ is shown. It is worthy to notice
that contrary to the sequential appearance of \textit{communicators} for the
Watts-Strogatz networks with $p\neq 0$, all units appear to be \textit{%
communicators} simultaneously for regular lattices $\left( p=0\right) $
owing to the indistinguishability between the units of the system.

\ \ 

$\left[ \text{Figure Caption}\right] $Fig.1: $\left( a\right) $ The first
appearance time of a \textit{communicator}, $t_{c}$\textit{\ }(the upper
part), and the corresponding degree of connection, $k$ (the lower part), for
different units of the system, $n$, where the unit $n$ is labelled in
accordance with the order of the $t_{c}$ value from small to large. $\left(
b\right) $ The average value of the first appearance times of the \textit{%
communicators}, $\left\langle t_{c}\right\rangle $, as a function of $k$ for
the results shown in $\left( a\right) $. \ \ 

The identification of \textit{communicators} may provide a powerful tool for
social dynamics. Here, we give an example by considering a merger between
two systems. Suppose that two systems, $P$ and $Q$, evolve according to the
dynamics of Eq. (\ref{eqa-1}) with the transition matrices $M_{P}$ and $%
M_{Q} $ which have the dimensions $N_{P}$ and $N_{Q}$. We further assume
that both $P$ and $Q$ are able to achieve some states of consensus, the 
\textit{theorem} then implies that there exists integers $n_{p}$ and $n_{q}$
such that $M_{P}^{n_{p}}$ and $M_{Q}^{n_{q}}$ have a positive column
locating respectively at, say, the $\alpha $th and the $\beta $th. As the
two systems are merged to form the system $R=P\cup Q$ by adding some
connections between $P$ and $Q$, the resultant transition matrix $M_{R}$
takes the form of 
\begin{equation}
M_{R}=\left( 
\begin{array}{cc}
M_{P} & C \\ 
C^{\tau } & M_{Q}%
\end{array}%
\right) ,  \label{eqa-15}
\end{equation}%
where the matrix $C$ specifies the connections between $P$ and $Q$, and $%
C^{\tau }$ is the transpose of $C$. Note that because of the added
connections, some entries of $M_{P}$ and $M_{Q}$ may have to be modified;
but, the positive entries remain to be positive after the modification, it
does not affect the result obtained below. By defining $n_{r}=\left[
n_{p},n_{q}\right] $, the least common multiple of $n_{p}$ and $n_{q}$, and
setting $\widehat{M}_{P}=\left( M_{P}^{n_{p}}\right) ^{n_{r}/n_{p}}$ and $%
\widehat{M}_{Q}=\left( M_{Q}^{n_{q}}\right) ^{n_{r}/n_{q}}$, we then have 
\begin{equation}
M_{R}^{n_{r}+1}\geq \left( 
\begin{array}{cc}
\widehat{M}_{P} & 0 \\ 
0 & \widehat{M}_{Q}%
\end{array}%
\right) \cdot M_{R},  \label{eqa-16}
\end{equation}%
which yields 
\begin{equation}
M_{R}^{n_{r}+1}\geq \left( 
\begin{array}{cc}
\widehat{M}_{P}\cdot M_{P} & \widehat{M}_{P}\cdot C \\ 
\widehat{M}_{Q}\cdot C^{\tau } & \widehat{M}_{Q}\cdot M_{Q}%
\end{array}%
\right) .  \label{eqa-16a}
\end{equation}%
Suppose that the new connection is added between the unit $\alpha $ of $P$
and the unit $\beta $ of $Q$, this gives a positive entry $\left( \alpha
,\beta \right) $ of $C$. Since the $\alpha $th column of $\widehat{M}_{P}$
and the $\beta $th column of $\widehat{M}_{Q}$ are positive, we have the $%
\beta $th column of $\widehat{M}_{P}\cdot C$ and the $\alpha $th column of $%
\widehat{M}_{Q}\cdot C^{\tau }$ being positive. Hence, the $\alpha $th and
the $\left( N_{P}+\beta \right) $th column of $M_{R}^{n_{r}+1}$ are
positive, and a state of consensus for the merged system $R$ can be achieved
according to the \textit{Theorem}. This gives the conclusion that only one
connection between two \textit{communicators} of different systems is
required for the existence of a state of consensus in the merged system.
However, the efficiency of reaching a consensus for the merged system
depends on the levels of the connected \textit{communicators} of different
systems. To show the dependence explicitly, we consider the mergence of two
systems defined in the Watts-Strogatz networks with $N=100$, $k_{0}=4$ for $%
p=0$ and $0.1$. The previous model for the transition matrix with $s=0.3$ is
used to classify the \textit{communicators} of two systems. All units are 
\textit{the primary communicators} for the system $p=0$, and the numbers of 
\textit{communicators} at different levels are different for the system $%
p=0.1$. By connecting a fixed unit of the system $p=0$ to one of the \textit{%
communicators} at a given level in the system $p=0.1$, we use the enlarged
transition matrix to calculate the time-steps of reaching the state of
consensus from a strongly disorder state, and then calculate the average
value over the time-steps required for different \textit{communicators} at
the same level in the system $p=0.1$. The results are shown as the plot of
the number of the average time-steps of reaching the state of consensus $%
c=0.5$, denoted as $\left\langle T\right\rangle $, vs. the level of the
connected \textit{communicator} of the system $p=0.1$, denoted as $L$, in
Fig. 2. Our results indicate that the connection between a pair of \textit{%
primary commutators} belonging to different systems provides the minimal and
the most efficient way to have the merged system reaching the state of
consensus. \ 

In summary, we present a novel way for characterizing the process of
reaching the state of consensus in a social system. The characterization
provide not only the insights on the occurrence of system-wide harmonic
behaviors but also a useful tool for the study of social dynamics. The
foundation for the characterization is the \textit{theorem} we establish,
which can be viewed as an important extension of the Perron-Frobenius
theorem.

\label{s5}

\label{s5-1}

\textbf{Acknowledgement}: We thank Yu-tin Huang for many stimulated
discussions. This work was partially supported by the National Science
Council of Republic of China (Taiwan) under the Grant No. NSC
99-2112-M-033-006 (MCH) and 99-2112-M-150-002 (YPL).

\FRAME{ftbpFU}{4.0041in}{1.638in}{0pt}{\Qcb{The first appearance time of
being a \textit{communicator}, $t_{c}$ (the upper part), and the
corresponding degrees of connection, $k$ (the lower part), for different
units of the system, $n$. Here, the unit $n$ is labelled in accordance with
the order of $t_{c}$ from small to large, and the system is defined on a
Watts-Strogatz network with $N=1000$, $k_{0}=4$, and $p=0.1$. }}{}{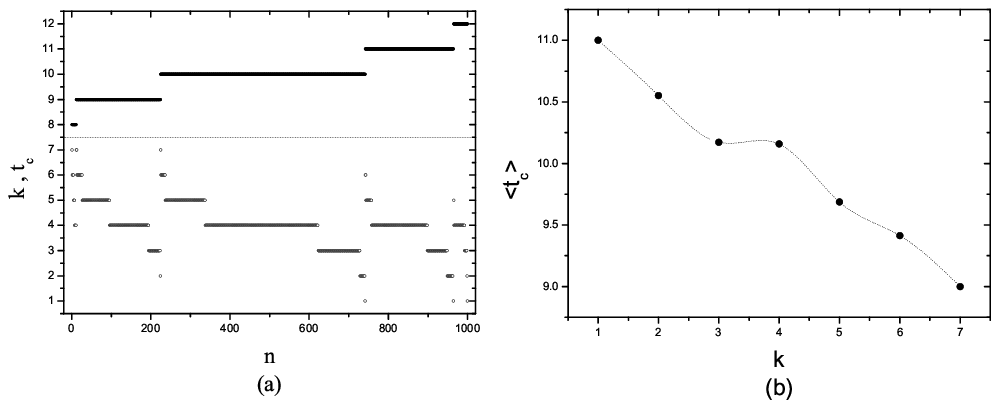}{%
\special{language "Scientific Word";type "GRAPHIC";maintain-aspect-ratio
TRUE;display "USEDEF";valid_file "F";width 4.0041in;height 1.638in;depth
0pt;original-width 3.9548in;original-height 1.6016in;cropleft "0";croptop
"1";cropright "1";cropbottom "0";filename 'fig1.ps';file-properties "XNPEU";}%
}\FRAME{ftbpFU}{2.399in}{1.8931in}{0pt}{\Qcb{The average of the first
appearance times, $\left\langle t_{c}\right\rangle $, of the \textit{%
communicators} with the same degree of connection, $k$, as a function of $k$%
. The results are based on the data of Fig. 1.}}{}{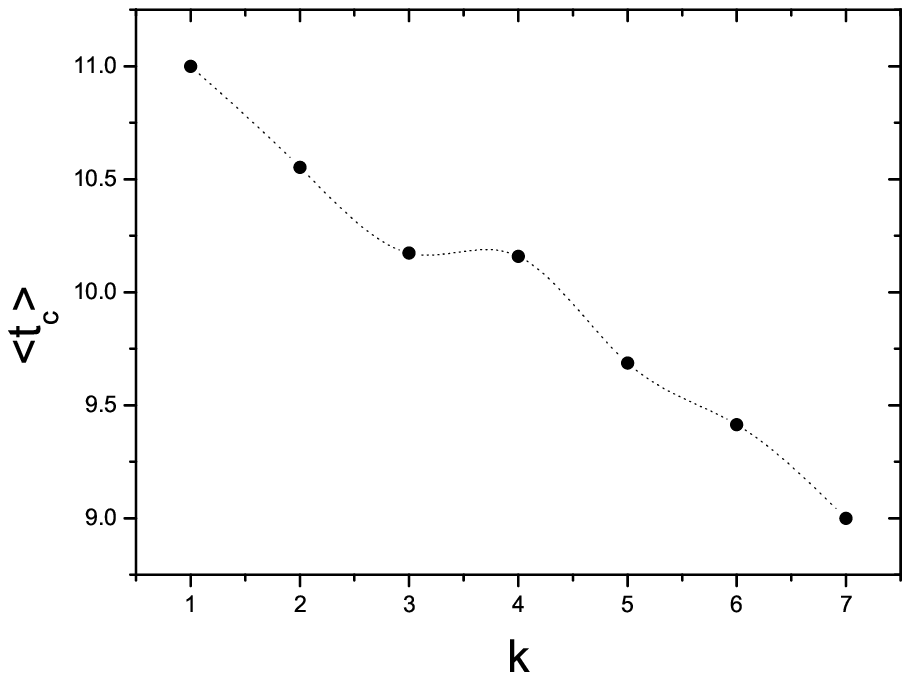}{\special%
{language "Scientific Word";type "GRAPHIC";maintain-aspect-ratio
TRUE;display "USEDEF";valid_file "F";width 2.399in;height 1.8931in;depth
0pt;original-width 4.2756in;original-height 3.3684in;cropleft "0";croptop
"1";cropright "1";cropbottom "0";filename 'fig2.eps';file-properties
"XNPEU";}}\FRAME{ftbpFU}{2.5789in}{2.0289in}{0pt}{\Qcb{The average
time-steps of reaching the consensus for the merged system, $\left\langle
T\right\rangle $, as a function of the level of the \textit{communicators}
of the second system, $L$, connected by a \textit{primary communicator} of
the first system. Two systems are defined on the Watts-Strogatz networks
with $N=100$, $k_{0}=4$, and the rewiring probability $p=0$ for the first
system and $p=0.1$ for the second. \ \ }}{}{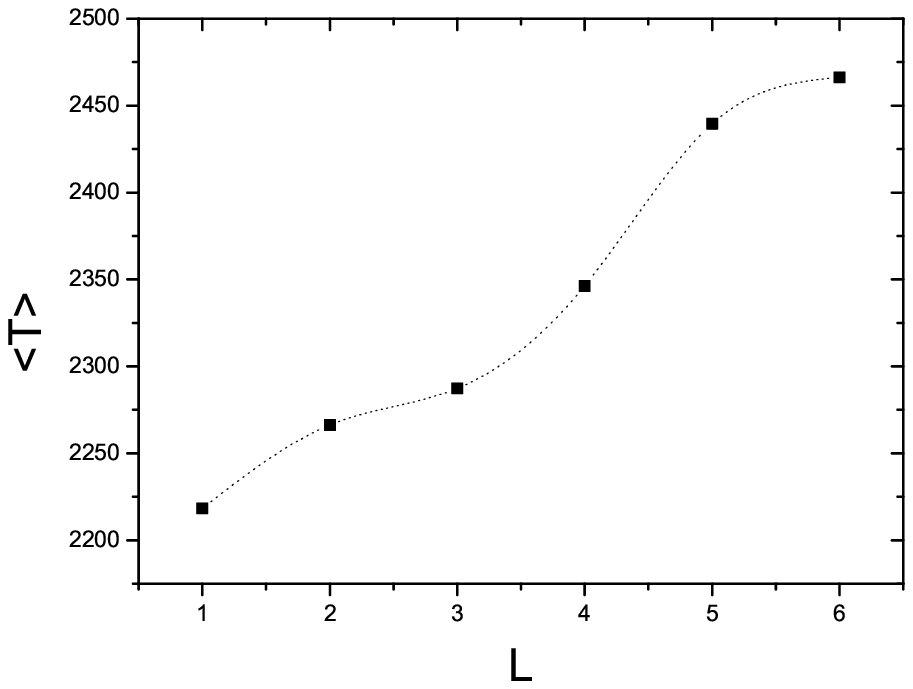}{\special{language
"Scientific Word";type "GRAPHIC";maintain-aspect-ratio TRUE;display
"USEDEF";valid_file "F";width 2.5789in;height 2.0289in;depth
0pt;original-width 4.2886in;original-height 3.3684in;cropleft "0";croptop
"1";cropright "1";cropbottom "0";filename 'fig3.eps';file-properties
"XNPEU";}}

\end{document}